\documentstyle[12pt,epsf]{article}
\newcommand{\be}{\begin{equation}}
\newcommand{\ee}{\end{equation}}
\newcommand{\bea}{\begin{eqnarray}}
\newcommand{\eea}{\end{eqnarray}}
\newcommand{\brr}{\begin{array}}
\newcommand{\err}{\end{array}}
\newcommand{\bc}{\begin{center}}
\newcommand{\ec}{\end{center}}
\newcommand{\nn}{\nonumber}
\newcommand{\as}{\alpha_{ s}}

\newcommand{\ep}{\epsilon}
\newcommand{\epp}{\epsilon^{\prime}}
\newcommand{\epse}{\epsilon^{\prime}/\epsilon}
\newcommand{\GF}{\frac{G_{ F}}{\sqrt 2}}
\newcommand{\Heff}{{\cal H}_{eff}}
\newcommand{\HH}{{\cal H}}
\newcommand{\DSone}{\Delta S\!=\!1}
\newcommand{\DStwo}{\Delta S\!=\!2}
\newcommand{\DBone}{\Delta B\!=\!1}
\newcommand{\DBtwo}{\Delta B\!=\!2}
\newcommand{\gammazeros}{\hat{\gamma}^{(0)T}_{ s}}

\newcommand{\W}{\hat{W}}

\newcommand{\alphas}{\alpha_{ s}}
\newcommand{\alphae}{\alpha_{ e}}

\newcommand{\Mw}{M_{ W}}
\begin{document}
\vskip 2truecm
{\Large
\centerline{\bf Estimates of $\epse$}
}
\vskip 1cm
\centerline{\bf M. Ciuchini$^{a}$, E. Franco$^b$, G. Martinelli$^{c}$
and L. Reina$^d$}

{\small
\centerline{$^a$ INFN, Sezione Sanit\`a, V.le Regina Elena 299,
00161 Rome, Italy. }
\centerline{$^b$ Dip. di Fisica,
Universit\`a degli Studi di Roma ``La Sapienza" and}
\centerline{INFN, Sezione di Roma, P.le A. Moro 2, 00185 Rome, Italy. }
\centerline{$^c$ Theory Division, CERN, 1211 Geneve 23, Switzerland.}
\centerline{$^d$ Brookhaven  National Lab., Physics Department, Upton,
NY 11973, U.S.A.}
}
\vskip 1cm
\begin{abstract}
We review the latest calculations of the next-to-leading $\DSone$ effective
Hamiltonian,
relevant for $K\to \pi\pi$ transitions. Numerical results for the Wilson
coefficients are given for different regularization schemes.
Predictions of $\epse$, obtained using different approaches to
evaluate the relevant hadronic matrix elements, are compared.
Given the present value of the top mass, $m_t=(174\pm 17)$ GeV,
all the analyses, in spite of the large theoretical uncertainties,
indicate that the value of $\epse$ is smaller than $1\times 10^{-3}$.
\end{abstract}
\vskip 1cm
\section{Introduction}
\label{sec:intro}

The understanding of mixing and CP-violation in hadronic systems is one
of the crucial tests of the Standard Model. In the last few years considerable
theoretical and experimental effort has been invested in this subject.

On the theoretical side, the complete next-to-leading expressions
of the relevant effective $\DSone$, $\DStwo$, $\DBone$
and $\DBtwo$ Hamiltonians have been computed \cite{altarelli}--\cite{cfmr1},
thus reducing the theoretical uncertainties\footnote{Indeed only
the top contribution to
the $\DStwo$ Hamiltonian is fully known, at the next-to-leading order.
To our knowledge, the other contributions have been only
partially computed \cite{hn}.}. Moreover, there is now
increasing theoretical evidence that the value of the
pseudoscalar $B$-meson decay constant is large, $f_B \sim 200$ MeV, and that
the $B$--$\bar B$ parameter $B_B$ is quite close to one.
This strongly constrains the Cabibbo-Kobayashi-Maskawa parameters and it
has remarkable consequences on CP-violation in $B$ decays,
see refs. \cite{reina,ciuc1}.
Still, the evaluation of hadronic matrix elements is subject to large
uncertainties, that are particularly severe for $\epse$, where
important cancellations of different contributions occur
for large values of the top mass. Indeed, a significative reduction of the
theoretical uncertainty on $\epse$ would require a substantial improvement in
the calculation of the hadronic matrix elements, either from
lattice simulations or from other non-perturbative techniques.

On the experimental side, more accurate measurements of the
mixing angles are now available and the mass of the top quark,
experimental evidence of which has recently been found by CDF \cite{top}, is
constrained within tight limits \cite{leptop}.
Still, in spite
of very accurate measurements, the experimental results for the $CP$
violating parameter $\epse$ are far from conclusive \cite{na31,e731}.
A better accuracy, at the level of $1\times 10^{-4}$, should be achieved
by the experiments of the next generation.

In the following, we briefly introduce the $\DSone$ effective Hamiltonian
and summarize the main results of the next-to-leading calculation of the
relevant Wilson coefficients.
An updated analysis of $\epse$, along the lines followed
in refs. \cite{reina}--\cite{epenew}, is presented. Particular emphasis is
devoted to
a realistic evaluation of the uncertainties. We also compare the results of
refs. \cite{reina}--\cite{epenew} to
the next-to-leading order analysis of ref. \cite{burasepe}.
In section \ref{sec:cp}, the basic formulae, which define the CP-violation
parameters $\ep$ and $\epp$, are presented;
in section \ref{sec:ckm}, the definition of the
Cabibbo--Kobayashi--Maskawa matrix and the notation used in this work are
introduced;
in sections \ref{sec:bare} and \ref{sec:effham}, we give several details about
the $\DSone$ effective Hamiltonian relevant for direct CP-violation. In
particular, the  Wilson coefficients in different regularization schemes are
reported.
In section \ref{sec:formule}, we give the formulae which has been used to
obtain
the theoretical  predictions; in section \ref{sec:results}, the
theoretical predictions from the more recent analyses are given.
Further details, including a theoretical discussion of the
matching conditions, of the $B$-parameters and of the uncertainties coming
from the choice of $\Lambda_{QCD}$, the renormalization scale, etc. can be
found in ref. \cite{epenew}.

\section{CP-violation in $K\to \pi\pi$ decays}
\label{sec:cp}

In this section, we introduce the parameters
$\ep$ and $\epp$  that describe CP-violation in the neutral kaon-system.
In the following, we assume CPT symmetry. A comprehensive
discussion of the general case, including CPT-violation, can be
found in ref. \cite{maiani0}.

There are two possible sources of CP-violation in the decays of the
neutral kaons into two pions. CP-violation can take place
both in the kaon mixing matrix and at the decay vertices.
Let us consider the mixing first.
The most general CPT-conserving Hamiltonian of the $K^0$--$\bar K^0$ system
at rest can be written as
\be
H=M-\frac{i}{2}\Gamma=\left(\brr{cc} M_0 & M_{12}\\M_{12}^* & M_0\err\right)
-\frac{i}{2}\left(\brr{cc} \Gamma_0 & \Gamma_{12}\\
\Gamma_{12}^* &\Gamma_{0}\err\right)\, ,
\ee
where the bra (ket) can be represented as the  two component vector
$\langle K^0 \vert\equiv (1,0)$, $\langle \bar K^0 \vert\equiv (0,1)$,
$M$ is the ``mass'' matrix and $\Gamma$ is the ``width'' matrix.
Both $M_0$ and $\Gamma_0$ are real.

Notice that there is some freedom in the definition of the
phases of the kaon field. In particular, one can make the change
\be
\vert K^0 \rangle \rightarrow e^{i\alpha}\vert K^0 \rangle\, , \qquad
\vert \bar K^0 \rangle \rightarrow e^{-i\alpha}\vert \bar K^0 \rangle\, .
\label{eq:fasech}
\ee
Correspondingly, the off--diagonal matrix elements of any operator $X$,
acting on the $K^0$--$\bar K^0$ system, undergo the  changes
\be
X_{12} \rightarrow e^{-2i\alpha}X_{12}\, ,\qquad
X_{21} \rightarrow e^{2i\alpha}X_{21}\, .
\ee
This arbitrariness enters in some popular
definitions of the CP-violation parameters.
For definiteness, we choose a particular phase convention,
namely we require that the CP operator is given by
\be
\mbox{CP}=\left(\brr{cc} 0 & 1\\1 & 0\err\right)\, .
\label{eq:fasedef}
\ee
In this case,
\be
\vert K_\pm\rangle =\frac{1}{\sqrt{2}}\left(\vert K^0\rangle \pm\vert
\bar K^0\rangle \right)
\ee
are the CP eigenstates.
In the presence of CP-violation, $\left[\mbox{CP},H\right]\ne 0$ and the
eigenstates of the Hamiltonian $H$
are not CP eigenstates.
We introduce the parameter $\bar\ep$ which defines the
eigenstates of $H$ as
\bea
\vert K_L\rangle =\frac{\vert K_-\rangle +\bar\ep \vert K_+\rangle }
{\sqrt{1+\vert \bar\ep\vert^2}}\, , \,\,\,\,\,\,\,\,\,
\vert K_S\rangle = \frac{\vert K_+\rangle +\bar\ep \vert K_-\rangle }
{\sqrt{1+\vert \bar\ep\vert^2}}\, .
\label{eq:epbar1}
\eea
The corresponding (complex) eigenvalues are denoted as
\bea
\lambda_L&=&m_L-i\Gamma_L/2=\frac{1}{2}\left(H_{11}+H_{22}\right)-
\sqrt{H_{12}H_{21}}\nn\\   \\
\lambda_S&=&m_S-i\Gamma_S/2=\frac{1}{2}\left(H_{11}+H_{22}\right)+
\sqrt{H_{12}H_{21}}\, . \nn
\eea
In the phase convention (\ref{eq:fasedef}), the parameter $\bar\ep$
controls the amount of CP-violation, namely the CP symmetric limit is recovered
for $\bar\ep\to 0$. We can explicitly write $\bar\ep$
in terms of the matrix elements of $H$
\be
\bar\ep=\frac{H_{21}-H_{12}}{\Delta\lambda-H_{12}-H_{21}}\, ,
\label{eq:epbar}
\ee
where $\Delta\lambda=\lambda_L-\lambda_S$.

Experimentally CP-violation is a small effect,
i.e. $\vert \bar\epsilon\vert \ll 1$. For this reason, one
can simplify eq. (\ref{eq:epbar}) to obtain
\be
\bar\ep\simeq-\frac{i\mbox{Im}M_{12}+\mbox{Im}\Gamma_{12}/2}{\Delta\lambda}\, ,
\label{eq:epsbardef}
\ee
with
\bea
\Delta\lambda&=&\Delta M-i\Delta\Gamma/2\nn\\
\Delta M&=&-2\mbox{Re}M_{12} \\
\Delta\Gamma&=&-2\mbox{Re}\Gamma_{12}\, . \nn
\eea
Moreover, since $\Delta M/\Delta\Gamma=-0.9565 \pm 0.0051 \approx -1$,
eq. (\ref{eq:epsbardef})
becomes
\be
\bar\ep\simeq\frac{1+i}{2}\frac{\mbox{Im}M_{12}}{2\mbox{Re}M_{12}}-\frac{1-i}{2}
\frac{\mbox{Im}\Gamma_{12}}{2\mbox{Re}\Gamma_{12}}\, .
\label{eq:epsbardef1}
\ee

In view of the following discussion of the CP-violation parameters,
let us introduce amplitudes of the weak decays
 of kaons into two pions states with definite isospin
\be
A_Ie^{i\delta_I}=\langle \pi\pi (I)\vert H_W\vert K^0\rangle\, ,
\ee
where $I=0,2$ is the isospin of the final two-pion state and
the $\delta_I$'s are the strong phases induced by final-state interaction.
Watson's theorem ensures that
\be
A_I^*e^{i\delta_I}=\langle \pi\pi (I)\vert H_W\vert \bar K^0\rangle\, .
\ee
Direct CP-violation, occurring at the decay vertices,
appears as a difference between the amplitudes
$\langle \pi\pi\vert H_W\vert K^0\rangle $ and
$\langle \pi\pi\vert H_W\vert\bar K^0\rangle $.
This corresponds to a phase difference between $A_0$ and
$A_2$.

One introduces the parameter $\epp$ to account for direct CP-violation.
A convenient definition is
\bea
\epp&=&
\frac{\langle \pi\pi(0)\vert H_W\vert K_S\rangle\langle\pi\pi (2)\vert
H_W\vert K_L\rangle-\langle \pi\pi(0)\vert H_W\vert K_L\rangle
\langle \pi\pi(2)\vert H_W\vert K_S\rangle}
{\sqrt{2}\langle \pi\pi(0)\vert H_W\vert K_S\rangle^2}\nn\\
&\simeq&i\frac{e^{i(\delta_2-\delta_0)}}{\sqrt{2}}\mbox{Im}(\frac{A_2}{A_0})
\simeq i\frac{e^{i(\delta_2-\delta_0)}}{\sqrt{2}}\frac{\omega}{\mbox{Re}A_0}
\left(\omega^{-1}\mbox{Im}A_2-\mbox{Im}A_0\right)\, ,
\label{eq:epspdef}
\eea
where $\omega=\mbox{Re}A_2/\mbox{Re}A_0$.
Equation (\ref{eq:epspdef}) is obtained in
the approximation
$\mbox{Im}A_0\ll \mbox{Re}A_0$,
$\mbox{Im}A_2\ll \mbox{Re}A_2$
and also $\omega\ll 1$, as a consequence of the $\Delta I=1/2$
enhancement  in  kaon decays;  $\epp$ is independent of the kaon phase
convention.
On the contrary,
the parameter $\bar\ep$, defined in eq. (\ref{eq:epbar1}),
depends on the choice of the phase.
Under a redefinition of the phases as in eq. (\ref{eq:fasech}),
$\bar\ep$ changes as
\be
\bar\ep\rightarrow\frac{-i\sin\alpha+\bar\ep\cos\alpha}
{\cos\alpha-i\bar\ep\sin\alpha}
\ee
and the CP-symmetric limit does not correspond to
$\bar\ep\to 0$\footnote{In this case, the states $\vert K_\pm\rangle$ are
not CP eigenstates.}.

Another   parameter, which is independent of the
phase convention and  accounts for  CP-violation in the mixing
matrix, can be defined in terms of the $K \to \pi \pi$
 transition amplitudes
\be
\ep =\frac{\langle \pi\pi (0)\vert H_W\vert K_L\rangle }
          {\langle \pi\pi (0)\vert H_W\vert K_S\rangle }
    =\frac{i\sin\phi_0+\bar\ep\cos\phi_0}{\cos\phi_0+i\bar\ep\sin\phi_0}
    \simeq\bar\ep+i\frac{\mbox{Im}A_0}{\mbox{Re}A_0}\, ,
\label{eq:epsdef}
\ee
where $A_0=\vert A_0 \vert e^{i\phi_0}$ and the last expression
is obtained in the approximation $\bar\ep,\,\phi_0 \ll 1$.
The two definitions, eqs. (\ref{eq:epbar1}) and (\ref{eq:epsdef}),
coincide in the Wu-Yang phase convention, $\mbox{Im}A_0=0$.
One can check that $\phi_0$ changes with the phase convention as
$\phi_0\rightarrow\phi_0+\alpha$ and that $\ep$ is invariant.

{}From unitarity, one has
\be
\Gamma_{12}=\sum_n 2\pi\delta(M_K-E_n)
\langle K^0\vert H_W\vert n\rangle
\langle n\vert H_W\vert \bar K^0\rangle\, .
\ee
Given the dominance of $K^0\to \pi\pi\,(\mbox{0})$ decay, one
obtains the relation $\Gamma_{12}=(A_0^*)^2$. From eqs.
(\ref{eq:epsbardef1}) and (\ref{eq:epsdef}), one has
\be
\ep\simeq\frac{e^{i\pi/4}}{\sqrt{2}}\left(\frac{\mbox{Im}M_{12}}{2\mbox{Re}M_{12}}-\xi\right)\, ,
\label{eq:epsdef1}
\ee
where $\xi=-\mbox{Im}A_0/\mbox{Re}A_0$. In the Cabibbo--Kobayashi--Maskawa
phase convention,
the $\xi$ contribution is small and can be safely neglected.

To make contact with the experiments, one defines
the two amplitude ratios
\bea
\eta_{00}= \frac{\langle \pi^0\pi^0\vert H_W\vert K_L\rangle }
{\langle \pi^0\pi^0\vert H_W\vert K_S\rangle }\, , \qquad
\eta_{+-} = \frac{\langle \pi^+\pi^-\vert H_W\vert K_L\rangle }
{\langle \pi^+\pi^-\vert H_W\vert K_S\rangle }\, .
\eea
Neglecting small terms, one has
\bea
\eta_{00} \simeq\ep-2\epp \, , \qquad
\eta_{+-} \simeq \ep+\epp\, ,
\eea
namely
\bea
\vert\epsilon\vert^2 &\simeq& \vert\eta_{+-}\vert^2
\simeq \vert\eta_{00}\vert^2\, ,\nn\eea
\vskip 0.2 cm \bea
\mbox{Re}\left(\frac{\epp}{\epsilon}\right)&\simeq&\frac{1}{6}\left(1-
\frac{\vert\eta_{00}\vert^2}
{\vert\eta_{+-}\vert^2}\right)\, .
\label{eq:parexp}
\eea
Expressing $\eta_{00}$ and $\eta_{+-}$ in terms of the corresponding
widths
\bea
\vert\eta_{00}\vert^2&=&\frac{\Gamma (K_L\to \pi^0\pi^0)}
{\Gamma (K_S\to \pi^0\pi^0)}\nn\\
\vert\eta_{+-}\vert^2&=&\frac{\Gamma (K_L\to \pi^+\pi^-)}
{\Gamma (K_S\to \pi^+\pi^-)}\, ,
\eea
eq. (\ref{eq:parexp}) gives the
CP-violation parameters in terms of measurable quantities. Notice that
$\epse$ is approximately
real, since
experimentally $\delta_2-\delta_0\approx -\pi/4$.

\section{The CKM matrix}
\label{sec:ckm}

In the Standard Model, the basic quark charged-current
 interactions are described by the Lagrangian
\be
{\cal L}_{{\rm quark-W}}=\frac{g}{2\sqrt 2}
{\bar u_i}\gamma_{\mu}(1-\gamma_5)V_{ij}d_jW^{\mu}
\hskip 2pt +\hskip 2pt {\rm h.c.}\, ,
\ee
where $u_i$ are the charged 2/3 quarks ($u,\hskip 1pt c,\hskip 1pt t$), $d_j$
the charged $-1/3$ quarks ($d,\hskip 1pt s,\hskip 1pt b$) and $g$ is the
$SU(2)_L$ weak coupling constant ($G_F/\sqrt2=g^2/8M_W^2$,
where $G_F$ is the Fermi constant).
$V$ is the
unitary CKM matrix \cite{ckm}. A useful parametrization is \cite{maiani,pdg}
\bea
V&=&\pmatrix{
V_{ud}&V_{us}&V_{ub}\cr
V_{cd}&V_{cs}&V_{cb}\cr
V_{td}&V_{ts}&V_{tb}\cr}\label{eq:ckm}\\
&=&\pmatrix{C_{\theta}C_{\sigma}&S_{\theta}C_{\sigma}&S_{\sigma}e^{-i\delta}\cr
-S_{\theta}C_{\tau}-C_{\theta}S_{\sigma}S_{\tau}e^{i\delta}&
C_{\theta}C_{\tau}-S_{\theta}S_{\tau}S_\sigma e^{i\delta}&C_{\sigma}S_{\tau}\cr
S_{\theta}S_{\tau}-C_{\theta}C_{\tau}S_{\sigma}e^{i\delta}&
-C_{\theta}S_{\tau}-C_{\tau}S_{\theta}S_{\sigma}e^{i\delta}&
C_{\sigma}C_{\tau}}\, .\nn
\eea
In eq. (\ref{eq:ckm}), $\theta$, $\sigma$ and $\tau$ are quark mixing angles
(in particular, $\theta$ corresponds approximately to the Cabibbo angle);
$C_{\theta},\hskip 2pt S_{\theta}$, etc., mean
$\cos\theta,\hskip 2pt \sin\theta$, etc.; $\delta$ is
the CP-violating phase. Experimental determinations of $\vert V_{ud}\vert$,
$\vert V_{cb}\vert$ and $\vert V_{ub}\vert$ from $K$ and $B$ decays show that
there is a hierarchy in the mixing angles, so that the CKM matrix can be
empirically
expanded in powers of $\lambda=S_{\theta}\simeq 0.22$ \cite{wolf}.
Up to and including terms of order $\lambda^3$ ($\lambda^5$)
for the real (imaginary) part, $V$ is given by
\be
V=\pmatrix{1-\frac{\lambda^2}{2}&\lambda&A\lambda^3\left(\rho-i\eta\right)\cr
-\lambda \left( 1+A^2\lambda^4(\rho+i\eta)\right)
&1-\frac{\lambda^2}{2}&A\lambda^2\cr
A\lambda^3\left[ 1-\left(1-\frac{\lambda^2}{2}\right)
(\rho+i\eta)\right]&-A\lambda^2\left(1+\lambda^2(\rho +i\eta)\right)&1\cr}\, ,
\ee
where $S_{\tau}=A\lambda^2$ and
$S_{\sigma}e^{-i\delta}=A\lambda^3(\rho-i\eta)$. In this particular
(quark) phase convention, the imaginary part of the matrix appears
at order $\lambda^3$.\par
\begin{figure}   
\begin{center}
\epsfxsize=.6\textwidth
\leavevmode\epsffile{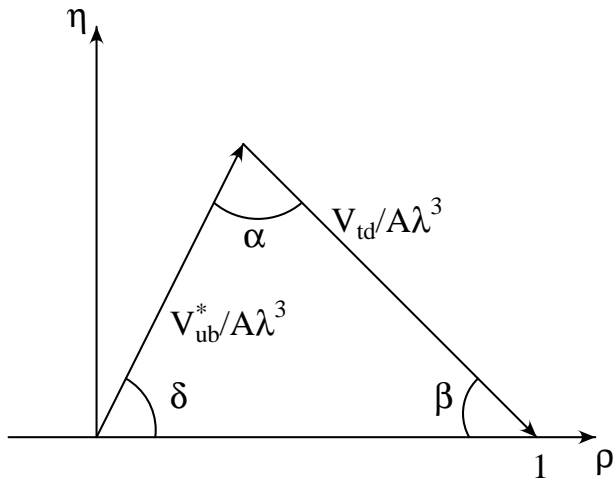}
\caption[]{\it The unitarity triangle in the $\rho$--$\eta$ plane.}
\label{fig:triangle}
\end{center}
\end{figure}

The unitarity of the CKM matrix implies
\be
\sum_q V_{iq}V_{jq}^*=\delta_{ij}\, .
\ee
In particular, considering the condition
\be
V_{ub}^*V_{ud}+V_{cb}^*V_{cd}+V_{tb}^*V_{td}=0
\label{eq:unitarity1}
\ee
in the approximation $V_{ud}\simeq V_{tb}\simeq 1$, one obtains
\be
\frac{V_{ub}^*}{A\lambda^3}+\frac{V_{td}}{A\lambda^3}-1=0\, .
\label{eq:unitarity2}
\ee
This relation identifies a triangle in the $\rho$--$\eta$ plane
(see fig. \ref{fig:triangle}).
The angles of this triangle, $\alpha$, $\beta$ and $\delta$,
are measures of CP-violation.

Recent phenomenological analyses of the CKM matrix elements can be found
in refs. \cite{epenew,alig,schubert}.
A brief discussion of these analyses together with
the  numerical results, can be found
in section \ref{sec:results}.

\section{The bare $\Delta S=1$ effective Hamiltonian}
\label{sec:bare}
Weak decays of light hadrons are more conveniently studied using the Wilson
operator
product expansion (OPE) \cite{wilson}.  With the OPE, it is possible to
introduce an
effective Hamiltonian, written in terms of renormalized local operators and of
the
corresponding Wilson coefficients \cite{bwlee}--\cite{gubpec}.
Short-distance strong-interaction effects
are contained in the coefficients and can be  computed  in perturbation theory,
because of
asymptotic freedom. Long-distance strong-interaction effects are included in
the hadronic
matrix elements of the local operators and must be evaluated with some
non-perturbative technique (lattice, QCD sum rules, etc.). The
convenience of the
effective Hamiltonian approach is that all
 known non-perturbative methods are usually
able to predict  matrix elements of local operators only.
In this section we introduce
the bare $\Delta S=1$ effective Hamiltonian, the renormalization of which
will be discussed in the next section.

At second order in the weak coupling constant and at zero order  in the
 strong coupling constant, the $\Delta S=1$ effective Hamiltonian can be
written in terms
of a local product of two charged currents
\bea \HH _W^{\Delta S=1}=
-\frac{G_F}{\sqrt{2}} \left[ \lambda_u \left( \bar
s_L \gamma_\mu u_L \right) \left( \bar
u_L \gamma_\mu d_L \right) + \lambda_c \left( u \to c \right)
\right]  =\nonumber \eea
\bea-\frac{G_F}{\sqrt{2}} \left[ \lambda_u \left( \bar
s_\alpha  u_\alpha \right)_{(V-A)} \left( \bar
u_\beta  d_\beta \right)_{(V-A)} + \lambda_c \left( u \to c \right)
\right]\, , \label{uno}
\eea
where $
\left({\bar s}_\alpha d_\alpha\right)_{(V-A)}=
{\bar s}_\alpha \gamma_{\mu}(1-\gamma_5)d_\alpha$;
 $\alpha$ and $\beta$ are colour indices
 and the sum over repeated indices is understood.
We have introduced the notation
\be \lambda_q=V_{qd}V^*_{qs} \ee for $q=u,c,t$.
In terms of the $\lambda_q$, the unitarity condition
of the CKM matrix can be written as
\be \lambda_u +\lambda_c + \lambda_t=0\, . \ee
Equation (\ref{uno})
 has been obtained from the original theory, by neglecting all
masses and momenta
with respect to  $M_W$.
In practice, the effective Hamiltonian
is obtained by taking $1/(M_W^2-q^2) \to 1/M_W^2$
 in the $T$-product of the two charged currents
and by putting   the $u$, $d$ and $s$ masses to zero.
In order to discuss CP-violation, it is convenient to
write $\HH _W^{\Delta S=1}$ as
\be \HH _W^{\Delta S=1}=
-\lambda_u \frac{G_F}{\sqrt{2}} \left[ \left(1-\tau \right)
\left( Q^u_2 -Q^c_2 \right) +\tau Q^u_2 \right] \, , \ee
where $\tau=-\lambda_t/\lambda_u$ contains the CP-violating phase and
\be Q^q_2=\left( \bar s_\alpha  q_\alpha \right)_{(V-A)} \left( \bar
q_\beta  d_\beta \right)_{(V-A)}\, . \ee
\section{QCD corrections}
\label{sec:effham}
Strong interactions play a crucial role in non-leptonic weak decays. The
perturbative
short-distance effects, included in the calculation of the Wilson coefficients,
may
be very important because of the presence of large logarithms $\sim \alpha_s^n
{\rm ln}^m (M_W/\mu)$, where $\mu$ is a scale of the order of the mass of the
decaying hadron.  For an accurate estimate of the short-distance
contributions, the large logarithms have to be resummed to all orders using
renormalization group (RG) techniques.

The starting point is the $T$-product of the two weak currents expanded at
short
distances in terms of local operators. Taking into account the
renormalization effects due to strong interactions, we write
\bea  \langle
F \vert  \HH _W^{\Delta S=1}\vert I \rangle=
 \frac{g^2}{8} \int d^4x D^{\mu \nu} \left( x, M_W \right) \langle
F \vert T \Bigl( J_\mu(x),J_\nu(0) \Bigr) \vert I \rangle = \nonumber \eea
\bea - \frac{G_F}{\sqrt{2}} \sum_i C_i (\mu) \langle F \vert Q_i(\mu) \vert I
\rangle
+ \dots \, , \label{due} \eea
where $\langle
F \vert$ and $\vert I \rangle$
are the generic final and initial states; the $Q_i(\mu)$ form a complete basis
of operators renormalized at the scale $\mu$;
 the $C_i(\mu)$ are the corresponding
Wilson coefficients and
the dots represent terms which are suppressed with respect to the dominant ones
as powers of $\Lambda_{{\rm QCD}}^2/ M_W^2$ ($m_b^2/M_W^2$ for $B$-decays).
The effective  Hamiltonian is
independent of renormalization scale $\mu$. On the
lattice, the renormalization scale can be replaced by the inverse
lattice spacing
$a^{-1}$ and the effective Hamiltonian can be expressed in terms of bare
 lattice operators \cite{epenew}. The OPE in eq. (\ref{due})
 must be valid for all possible initial and final
states. This implies that
 the effective Hamiltonian is defined from an operator relation
\bea \HH _W^{\Delta S=1}=-\frac{G_F}{\sqrt{2}}  \sum_i C_i (\mu)  Q_i(\mu)=
-\frac{G_F}{\sqrt{2}} \vec Q^T(\mu) \cdot \vec C(\mu) \, .\label{effH} \eea

The important features of $\HH _W^{\Delta S=1}$ are the following:
\begin{itemize}
\item the Wilson coefficients can
be calculated using (RG-improved) perturbation theory, provided that one
chooses a sufficiently large renormalization scale $\mu \simeq 2$--$3$ GeV
$\gg \Lambda_{{\rm QCD}}$.
In the leading logarithmic approximation (LLA),
all  terms of  $O\left(\alpha_s(\mu)^n\log
(\Mw/\mu)^n\right)$ are taken into account;

\item all non-perturbative effects are contained  in the matrix elements of the
local operators, the calculation of which requires a non-perturbative
technique.
\end{itemize}

Since  $\HH _W^{\Delta S=1}$,  eq. (\ref{effH}), is
independent of $\mu$, the coefficients $\vec C(\mu)=\left(C_1(\mu), C_2(\mu),
\dots\right)$ must satisfy the RG equations
\be
\mu^2\frac{d}{d\mu^2}\vec C(\mu)=\frac{1}{2}\hat\gamma^T\vec C(\mu)\, ,
\ee
which can be more  conveniently written as
\be
\left(\mu^2\frac{\partial}{\partial\mu^2}+\beta(\as)
\frac{\partial}{\partial\as}-\frac{1}{2}\hat\gamma^T(\as)\right)
\vec C(\mu)=0\, ,
\label{eq:rge}
\ee
where
\be
\beta(\alpha_s)=\mu^2\frac{d\alpha_s}{d\mu^2}
\ee
is the QCD $\beta$-function and
\be
\hat\gamma(\alpha_s)=2\hat Z^{-1}\mu^2\frac{d}{d\mu^2}\hat Z
\ee
is the anomalous-dimension matrix of the renormalized operators;
$\hat Z$ is defined by the relation which
connects the bare operators to the renormalized ones,
$\vec Q(\mu)=\hat Z^{-1}(\mu, \alpha_s) \vec Q^B$.

The solution of the system of linear  equations (\ref{eq:rge}) is
 found by introducing a suitable
evolution matrix $U(\mu,M_W)$
and by imposing an appropriate set of initial
conditions, usually called matching conditions.
The coefficients $\vec C(\mu)$ are given by\footnote{ The problem of the
thresholds due to the presence of heavy quarks with a mass $M_W \gg m_Q \gg
 \Lambda_{{\rm QCD}}$ will be discussed below.}
\be
\vec C(\mu)=\hat U(\mu,M_W)\vec C(M_W)\, ,
\label{eq:cmu}
\ee
with
\be
\hat U(m_1,m_2)=T_{\as}\exp\left(\int_{\as(m_1)}^{\as(m_2)}
\frac{d\as}{\beta(\as)}\hat\gamma^T(\as)\right) \, ;
\label{eq:solU}
\ee
 $T_{\as}$ is the ordered product with
increasing couplings from right to left.
The matching  conditions are found by imposing that, at $\mu=M_W$,
the matrix elements of the original $T$-product of the currents coincide,
 up to terms
suppressed as inverse powers of $M_W$, with  the corresponding
matrix elements of $\HH _W^{\Delta S=1}$. To this end,
we introduce the vector  $\vec T$ defined by the relation
\be
\langle \alpha\vert T (J^\dagger J)
\vert \beta \rangle =-\frac{G_F}{\sqrt{2}}\langle \alpha\vert
\vec Q^T \vert \beta \rangle_0 \cdot \vec T + \dots
\label{eq:matcfull}
\ee
where $\langle \alpha\vert
\vec Q^T \vert \beta \rangle _0$ are the matrix elements
of the operators at  tree level.
We also introduce the matrix $\hat M(\mu)$ such that
\bea \langle \alpha\vert \HH _W^{\Delta S=1}\vert \beta \rangle
=-\frac{G_F}{\sqrt{2}} \langle \alpha\vert
\vec Q^T(\mu) \vert \beta \rangle \vec C(\mu) = \nonumber \eea \bea
-\frac{G_F}{\sqrt{2}}\langle \alpha\vert
\vec Q^T \vert \beta \rangle_0\hat M^T(\mu)\vec C(\mu)\, .
\label{eq:matceff}
\eea
In terms of $\vec T$ and $\hat M$, the matching condition
\be
\langle \alpha\vert T (J^\dagger J) \vert \beta \rangle=
\langle \alpha\vert \HH _W^{\Delta S=1}\vert \beta \rangle \,
\ee
fixes  the value of the Wilson coefficients at the scale $M_W$
\be
\vec C(M_W)=[\hat M^T(M_W)]^{-1}\vec T\, .
\label{eq:inicond}
\ee
Notice that the matching could be imposed at any scale $\bar\mu$,
such that large logarithms do not appear
in the calculation of the Wilson coefficients at the
scale $\bar \mu$, i.e.  $\as \ln M_W/\bar\mu
\ll 1$.

Equation  (\ref{eq:cmu}) is correct
if no  threshold corresponding to a quark mass
 between $\mu$ and $M_W$ is present. Indeed, as $\as$,
$\hat\gamma$ and $\beta(\as)$ all
 depend on the number of active flavours, it is necessary to change
the evolution matrix $\hat U$ defined in  eq.
(\ref{eq:solU}), when passing the threshold.
 The general case then corresponds
to  a sequence
of effective theories with a decreasing number of ``active'' flavours.
By ``active'' flavour, we mean a dynamical massless
($\mu \gg m_Q$) quark field. The
theory with $k$ ``active'' flavours is matched to the one with $k+1$
``active'' flavours at the threshold. This procedure changes the solution
for the Wilson coefficients. For instance, if one starts
with five ``active'' flavours at the scale $M_W$ and chooses $m_c
\ll \mu \ll m_b$,
 the Wilson coefficients become
\be
\vec C(\mu)= \W[\mu,M_W] \vec C(M_W)
= \hat U_4(\mu,m_b)\hat U_5(m_b,M_W)\vec C(M_W)\, .
\label{ric} \ee
The inclusion of the charm threshold proceeds along the same lines.
\section{The operators of the $\Delta S=1$ effective Hamiltonian}
\begin{figure}   
\begin{center}
\epsfxsize=.6\textwidth
\epsfysize=.9\epsfxsize
\leavevmode\epsffile{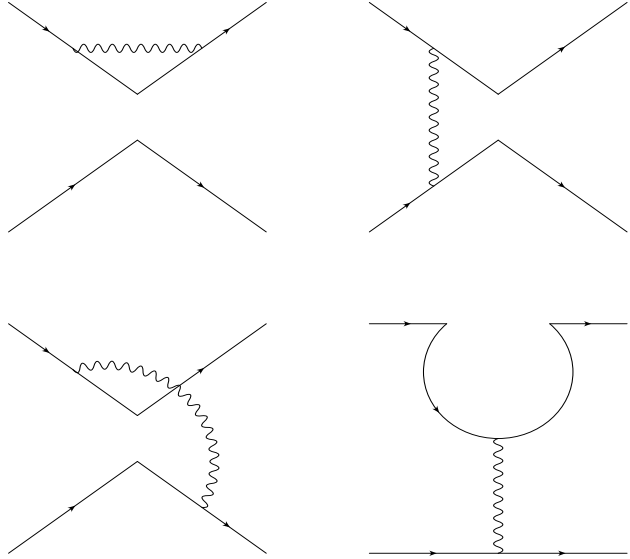}
\caption[]{\it One-loop corrections to the $\DSone$ effective Hamiltonian.}
\label{fig:oneloop}
\end{center}
\end{figure}

So far, we have presented the exact solutions
of the renormalization group equations for  the Wilson coefficients. In
practice, it is only possible  to calculate the relevant functions in
perturbation
theory. For illustrative purposes, we consider
the calculation of the $\DSone$ effective Hamiltonian at the leading order in
QCD. The bare Hamiltonian is given in eq. (\ref{uno}).
In the presence of QCD interactions, other operators appear
in the Wilson expansion.
A complete basis is given by the following operators
\bea
Q_{ 1}&=&({\bar s}_{\alpha}d_{\alpha})_{ (V-A)}
    ({\bar u}_{\beta}u_{\beta})_{(V-A)}
   \nn\\
Q_{ 2}&=&({\bar s}_{\alpha}d_{\beta})_{ (V-A)}
    ({\bar u}_{\beta}u_{\alpha})_{ (V-A)}
\nn \\
Q_{ 3} &=& ({\bar s}_{\alpha}d_{\alpha})_{ (V-A)}
    \sum_q({\bar q}_{\beta}q_{\beta})_{ (V- A)}
\nn \\
Q_{ 4} &=& ({\bar s}_{\alpha}d_{\beta})_{ (V-A)}
    \sum_q({\bar q}_{\beta}q_{\alpha})_{ (V- A)}
\nn \\
Q_{ 5} &=& ({\bar s}_{\alpha}d_{\alpha})_{ (V-A)}
    \sum_q({\bar q}_{\beta}q_{\beta})_{ (V+ A)}
\nn \\
Q_{ 6} &=& ({\bar s}_{\alpha}d_{\beta})_{ (V-A)}
    \sum_q({\bar q}_{\beta}q_{\alpha})_{ (V+ A)}
\nn \\
Q^c_{ 1}&=&({\bar s}_{\alpha}d_{\alpha})_{ (V-A)}
    ({\bar c}_{\beta}c_{\beta})_{ (V-A)}
\nn \\
Q^c_{ 2}&=&({\bar s}_{\alpha}d_{\beta})_{ (V-A)}
    ({\bar c}_{\beta}c_{\alpha})_{ (V-A)}\, .
\eea
The $q$ index  runs over the ``active'' flavours.
The above  operators are generated by gluon exchanges
in the Feynman diagrams of fig.
\ref{fig:oneloop}. In particular, $Q_1$ is generated by  current--current
diagrams and $Q_3$--$Q_6$ are generated by penguin diagrams. The
choice of the operator basis in not unique, and different possibilities have
been considered in the literature \cite{basis}. If the electromagnetic
correction, are also taken into account, the operator basis enlarges
to include the following operators
\bea
Q_{7} &=& \frac{3}{2}({\bar s}_{\alpha}d_{\alpha})_
    { (V-A)}\sum_{q}e_{ q}({\bar q}_{\beta}q_{\beta})_
    { (V+ A)}
\nn \\
Q_{8} &=& \frac{3}{2}({\bar s}_{\alpha}d_{\beta})_
    { (V-A)}\sum_{q}e_{ q}({\bar q}_{\beta}q_{\alpha})_
    { (V+ A)} \nn \\
\nn\\
Q_{9} &=& \frac{3}{2}({\bar s}_{\alpha}d_{\alpha})_
    { (V-A)}\sum_{q}e_{ q}({\bar q}_{\beta}q_{\beta})_
    { (V- A)}
\nn \\
Q_{10} &=& \frac{3}{2}({\bar s}_{\alpha}d_{\beta})_
    { (V-A)}\sum_{q}e_{ q}({\bar q}_{\beta}q_{\alpha})_
    { (V-A)}\, .
\eea
Below the bottom threshold, the following relation holds
\be
Q_{10}-Q_{9}-Q_4+Q_3=0\, ,
\label{eq:q10}
\ee
so that there are nine independent operators. The basis is further reduced
below the charm threshold by using the relations
\bea
Q_4-Q_3-Q_2+Q_1&=&0\nn\\
Q_9-\frac{3}{2}Q_1+\frac{1}{2}Q_3&=&0\, .
\eea

All the operators considered above
are dimension-six operators. In principle, two dimension-five
 operators
\bea
Q_{11}&=&\frac{Q_de}{16\pi^2}m_s\bar s_\alpha\sigma^{\mu\nu}_{(V-A)}d_\alpha
F_{\mu\nu}\nn\\
Q_{12}&=&\frac{g}{16\pi^2}m_s\bar s_\alpha\sigma^{\mu\nu}_{(V-A)}
t^A_{\alpha\beta}d_\beta G^A_{\mu\nu}
\eea
should also be included in the operator basis.
The matrix elements of $Q_{11}$ and
$Q_{12}$, however, enter only at  $O(p^4)$
in chiral perturbation theory. Since the phenomenological analysis
presented in the following is  only valid up
to terms of  $O(p^2)$, we do not need to include the contribution
of the dimension-five operators in the  calculation of $\epse$.
The effect of these operators on $\epse$  has  recently been
analysed in  ref. \cite{mimmo}.
Other operators of lower dimensionality
(e.g.  two-fermion operators) are also potentially present.
 However,   it can be shown
that  their effect can be reabsorbed  in  a suitable redefinition
of the fermion fields and by diagonalizing the quark mass matrix
at first order in  $G_F$ \cite{bwlee}--\cite{gubpec}.

In  summary, the $\DSone$ effective Hamiltonian, renormalized at a scale
$\mu \gg
m_c$, can be written as
\bea
{\cal H}_{eff}^{\DSone}&=&-\lambda_u \frac {G_F} {\sqrt{2}}
\Bigl\{ (1 - \tau ) \Bigl[ C_1(\mu)\left( Q_1(\mu) - Q_1^c(\mu) \right) +
C_2(\mu)\left( Q_2(\mu) - Q_2^c(\mu) \right)  \Bigr]\nn\\
& &+\tau \vec Q(\mu)^T \vec C(\mu) \Bigr\}\, ,
\eea
where, in order to find the Wilson coefficients to a given order
in $\as$, we have to calculate eqs.
(\ref{eq:solU}), (\ref{eq:inicond}) in perturbation theory.

The explicit expressions  of $\hat\gamma_s(\as)$ and $\beta(\as)$,
in the LLA
\bea
\hat \gamma(\as) = \frac {\alphas }{ 4 \pi } \hat \gamma_s^{(0)}\, , \qquad
\beta(\as) =-\frac{\as^2}{4\pi}\beta_0 \, ,
\eea
 can be found for example in  ref. \cite{cfmr1}.
In eq. (\ref{eq:solU}), using  $\hat\gamma^{(0)}_s$ and $\beta_0$,
 one obtains
\be
U(\mu,M_W)=\left(\frac{\alphas (M_W)}{\alphas (\mu)}\right)^
{\gammazeros / 2\beta_{ 0}}\, .
\ee

At this order, the matching conditions are trivial:
 $\hat M$, eq. (\ref{eq:matceff}), is the identity matrix;
$\vec T$, eq. (\ref{eq:matcfull}), has all vanishing components with
the only exception of  $T_2=1$.
 Thus the Wilson coefficients at the leading order for
$m_c \ll \mu \ll m_b$ are given by
\be
\vec C(\mu)=\left(\frac{\alphas^{n_f=4} (m_b)}{\alphas^{n_f=4} (\mu)}\right)^
{(\gammazeros / 2\beta_0)_{n_f=4}}
\left(\frac{\alphas^{n_f=5} (M_W)}{\alphas^{n_f=5} (m_b)}\right)^
{(\gammazeros / 2\beta_0)_{n_f=5}}\vec C(\Mw)\, ,
\ee
with  $C_2(\Mw)=1$ and all the other Wilson coefficients at the
scale $M_W$  vanish.

In the next-to-leading logarithmic approximation
(NLLA), one proceeds along the general scheme
described above.  In this case, all quantities entering in the
matching procedure have to be computed at order $\as$ ($\alpha_e$ for the
electromagnetic case). The $\beta$-function and the anomalous dimension
matrix have to be computed at second order in the coupling constants.
Thus, for example,
the anomalous dimension matrix in the NLLA has the form
\be
\hat \gamma= \frac {\alphas }{ 4 \pi } \hat \gamma_s^{(0)} +
 \frac {\alphae }{4 \pi} \hat \gamma_e^{(0)}
+ (\frac {\alphas }{4 \pi})^2 \hat \gamma_s^{(1)} +
 \frac{ \alphas }{4 \pi} \frac{ \alphae}{4 \pi}  \hat \gamma_e^{(1)}\, ,
\ee
where  $O(\alphae^2)$ corrections have been neglected.
We will not give here any details of the NLLA calculations. They
can be found in refs. \cite{altarelli}--\cite{cfmr1}.
At the next-to-leading order,
it is necessary to solve numerically eq.
 (\ref{eq:rge}).
Table \ref{tab:coeff} contains the coefficients,
calculated at the leading (LO) and at the next-to-leading (NLO) order,
using the 't Hooft--Veltman (HV) and the na\"\i ve dimensional (NDR)
regularization
schemes, for different values of the renormalization scale $\mu$.
The errors in the table take into account the variation of the
values of the coefficients due to $\Lambda_{QCD}^{(4)}=(330\pm 100)$ MeV
and $m_t=(174\pm 17)$ GeV.
Notice that the next-to-leading Wilson coefficients and operators both depend
on the regularization scheme, while the effective Hamiltonian is
scheme-independent
up to terms $O(\as^2)$. Actually the dependence of the effective
Hamiltonian on the regularization scheme, due to the unknown
next-to-next-to-leading terms, can be estimated and contributes to the
uncertainties in the prediction of $\epse$, see ref. \cite{epenew}.

The coefficients in table \ref{tab:coeff} have been computed independently by
the Munich group \cite{bur3,burasepe}.
The definition of the renormalized operators in the HV scheme used here
differ from those defined in ref. \cite{burasepe}.
This is due to the different way of taking into account the two-loop
anomalous dimension of the weak current, which does not vanish in the HV
calculation. One can relate the HV coefficients of table \ref{tab:coeff}
($\vec C$) and those of ref. \cite{burasepe} ($\vec C'$). The relation is
\be
\vec C(\mu)=\Bigl(1-\frac{\alphas(\mu)}{4\pi}\frac{\gamma_J}{\beta_0}
\hat 1\Bigr)\vec C'(\mu)\, ,
\ee
where
\be
\gamma_J=4\frac{N_c^2-1}{2N_c}\beta_0\, .
\ee
Once these differences in the definition of the renormalized operators and the
reduction of the operator basis, eq. (\ref{eq:q10}), are properly taken into
account, the numerical results presented here agree with
those of ref. \cite{burasepe}.
{
\scriptsize
\begin{table}
\begin{center}
\begin{tabular}{|c|c|c|c|}\hline
 & LO & NLO HV & NLO NDR\\\hline\hline
& \multicolumn{3}{c|}{$\mu=1.5$ GeV}\\\hline
$C_{1}$ & $(-4.22\pm 0.65\pm 0.00)\times 10^{-1}$
& $(-3.91\pm 0.51\pm 0.00)\times 10^{-1}$
& $(-3.80\pm 0.55\pm 0.00)\times 10^{-1}$
\\
$C_{2}$ & $(11.62\pm 0.38\pm 0.00)\times 10^{-1}$
& $(106.13\pm 0.82\pm 0.00)\times 10^{-2}$
& $(11.95\pm 0.35\pm 0.00)\times 10^{-1}$
\\
$C_{3}$ & $(1.99\pm 0.35\pm 0.00)\times 10^{-2}$
& $(2.17\pm 0.41\pm 0.00)\times 10^{-2}$
& $(2.60\pm 0.52\pm 0.00)\times 10^{-2}$
\\
$C_{4}$ & $(-4.16\pm 0.56\pm 0.02)\times 10^{-2}$
& $(-4.51\pm 0.60\pm 0.01)\times 10^{-2}$
& $(-0.63\pm 0.11\pm 0.00)\times 10^{-1}$
\\
$C_{5}$ & $(1.19\pm 0.12\pm 0.00)\times 10^{-2}$
& $(1.37\pm 0.15\pm 0.00)\times 10^{-2}$
& $(10.52\pm 0.61\pm 0.01)\times 10^{-3}$
\\
$C_{6}$ & $(-0.66\pm 0.13\pm 0.00)\times 10^{-1}$
& $(-0.63\pm 0.11\pm 0.00)\times 10^{-1}$
& $(-0.93\pm 0.21\pm 0.00)\times 10^{-1}$
\\
$C_{7}$ & $(0.16\pm 0.04\pm 0.19)\times 10^{-3}$
& $(-0.04\pm 0.00\pm 0.17)\times 10^{-3}$
& $(0.02\pm 0.06\pm 0.20)\times 10^{-3}$
\\
$C_{8}$ & $(0.63\pm 0.14\pm 0.16)\times 10^{-3}$
& $(0.97\pm 0.19\pm 0.15)\times 10^{-3}$
& $(1.06\pm 0.26\pm 0.19)\times 10^{-3}$
\\
$C_{9}$ & $(-6.77\pm 0.27\pm 0.71)\times 10^{-3}$
& $(-6.32\pm 0.37\pm 0.64)\times 10^{-3}$
& $(-7.24\pm 0.19\pm 0.73)\times 10^{-3}$
\\
\hline\hline
& \multicolumn{3}{c|}{$\mu=2$ GeV}\\\hline
$C_{1}$ & $(-3.47\pm 0.44\pm 0.00)\times 10^{-1}$
& $(-3.29\pm 0.37\pm 0.00)\times 10^{-1}$
& $(-3.13\pm 0.39\pm 0.00)\times 10^{-1}$
\\
$C_{2}$ & $(11.16\pm 0.23\pm 0.00)\times 10^{-1}$
& $(104.13\pm 0.54\pm 0.00)\times 10^{-2}$
& $(11.54\pm 0.23\pm 0.00)\times 10^{-1}$
\\
$C_{3}$ & $(1.59\pm 0.23\pm 0.00)\times 10^{-2}$
& $(1.73\pm 0.26\pm 0.00)\times 10^{-2}$
& $(2.07\pm 0.33\pm 0.00)\times 10^{-2}$
\\
$C_{4}$ & $(-3.50\pm 0.40\pm 0.01)\times 10^{-2}$
& $(-3.82\pm 0.44\pm 0.01)\times 10^{-2}$
& $(-5.19\pm 0.71\pm 0.01)\times 10^{-2}$
\\
$C_{5}$ & $(10.40\pm 0.94\pm 0.04)\times 10^{-3}$
& $(1.20\pm 0.11\pm 0.00)\times 10^{-2}$
& $(10.54\pm 0.16\pm 0.02)\times 10^{-3}$
\\
$C_{6}$ & $(-5.23\pm 0.80\pm 0.03)\times 10^{-2}$
& $(-5.08\pm 0.72\pm 0.03)\times 10^{-2}$
& $(-0.72\pm 0.13\pm 0.00)\times 10^{-1}$
\\
$C_{7}$ & $(0.18\pm 0.02\pm 0.19)\times 10^{-3}$
& $(0.01\pm 0.00\pm 0.18)\times 10^{-3}$
& $(0.01\pm 0.04\pm 0.20)\times 10^{-3}$
\\
$C_{8}$ & $(0.50\pm 0.08\pm 0.12)\times 10^{-3}$
& $(0.77\pm 0.12\pm 0.12)\times 10^{-3}$
& $(0.81\pm 0.16\pm 0.14)\times 10^{-3}$
\\
$C_{9}$ & $(-7.03\pm 0.20\pm 0.74)\times 10^{-3}$
& $(-6.71\pm 0.27\pm 0.68)\times 10^{-3}$
& $(-7.49\pm 0.15\pm 0.75)\times 10^{-3}$
\\
\hline\hline
& \multicolumn{3}{c|}{$\mu=3$ GeV}\\\hline
$C_{1}$ & $(-2.68\pm 0.28\pm 0.00)\times 10^{-1}$
& $(-2.59\pm 0.25\pm 0.00)\times 10^{-1}$
& $(-2.41\pm 0.25\pm 0.00)\times 10^{-1}$
\\
$C_{2}$ & $(10.71\pm 0.12\pm 0.00)\times 10^{-1}$
& $(101.88\pm 0.24\pm 0.00)\times 10^{-2}$
& $(11.12\pm 0.14\pm 0.00)\times 10^{-1}$
\\
$C_{3}$ & $(1.20\pm 0.14\pm 0.01)\times 10^{-2}$
& $(1.29\pm 0.16\pm 0.00)\times 10^{-2}$
& $(1.56\pm 0.19\pm 0.01)\times 10^{-2}$
\\
$C_{4}$ & $(-2.78\pm 0.26\pm 0.01)\times 10^{-2}$
& $(-3.06\pm 0.29\pm 0.01)\times 10^{-2}$
& $(-4.10\pm 0.46\pm 0.00)\times 10^{-2}$
\\
$C_{5}$ & $(8.60\pm 0.67\pm 0.04)\times 10^{-3}$
& $(9.99\pm 0.84\pm 0.04)\times 10^{-3}$
& $(9.75\pm 0.46\pm 0.03)\times 10^{-3}$
\\
$C_{6}$ & $(-3.89\pm 0.47\pm 0.03)\times 10^{-2}$
& $(-3.84\pm 0.44\pm 0.03)\times 10^{-2}$
& $(-5.33\pm 0.73\pm 0.03)\times 10^{-2}$
\\
$C_{7}$ & $(0.22\pm 0.01\pm 0.20)\times 10^{-3}$
& $(0.08\pm 0.00\pm 0.19)\times 10^{-3}$
& $(0.04\pm 0.02\pm 0.20)\times 10^{-3}$
\\
$C_{8}$ & $(3.73\pm 0.50\pm 0.92)\times 10^{-4}$
& $(5.84\pm 0.74\pm 0.91)\times 10^{-4}$
& $(0.59\pm 0.09\pm 0.10)\times 10^{-3}$
\\
$C_{9}$ & $(-7.29\pm 0.14\pm 0.77)\times 10^{-3}$
& $(-7.04\pm 0.18\pm 0.73)\times 10^{-3}$
& $(-7.66\pm 0.11\pm 0.78)\times 10^{-3}$
\\
\hline\hline
& \multicolumn{3}{c|}{$\mu=4.5$ GeV}\\\hline
$C_{1}$ & $(-2.07\pm 0.19\pm 0.00)\times 10^{-1}$
& $(-2.03\pm 0.18\pm 0.00)\times 10^{-1}$
& $(-1.84\pm 0.17\pm 0.00)\times 10^{-1}$
\\
$C_{2}$ & $(103.85\pm 0.67\pm 0.00)\times 10^{-2}$
& $(1001.15\pm 0.43\pm 0.00)\times 10^{-3}$
& $(108.20\pm 0.87\pm 0.00)\times 10^{-2}$
\\
$C_{3}$ & $(1.13\pm 0.11\pm 0.03)\times 10^{-2}$
& $(1.19\pm 0.12\pm 0.03)\times 10^{-2}$
& $(1.40\pm 0.14\pm 0.03)\times 10^{-2}$
\\
$C_{4}$ & $(-2.42\pm 0.20\pm 0.02)\times 10^{-2}$
& $(-2.67\pm 0.23\pm 0.02)\times 10^{-2}$
& $(-3.51\pm 0.34\pm 0.02)\times 10^{-2}$
\\
$C_{5}$ & $(7.06\pm 0.49\pm 0.04)\times 10^{-3}$
& $(8.36\pm 0.66\pm 0.04)\times 10^{-3}$
& $(8.75\pm 0.52\pm 0.04)\times 10^{-3}$
\\
$C_{6}$ & $(-2.94\pm 0.29\pm 0.02)\times 10^{-2}$
& $(-2.97\pm 0.29\pm 0.02)\times 10^{-2}$
& $(-4.09\pm 0.46\pm 0.02)\times 10^{-2}$
\\
$C_{7}$ & $(0.29\pm 0.01\pm 0.20)\times 10^{-3}$
& $(0.17\pm 0.00\pm 0.19)\times 10^{-3}$
& $(0.09\pm 0.01\pm 0.20)\times 10^{-3}$
\\
$C_{8}$ & $(2.88\pm 0.32\pm 0.69)\times 10^{-4}$
& $(4.53\pm 0.49\pm 0.70)\times 10^{-4}$
& $(4.39\pm 0.52\pm 0.77)\times 10^{-4}$
\\
$C_{9}$ & $(-0.96\pm 0.01\pm 0.10)\times 10^{-2}$
& $(-0.93\pm 0.00\pm 0.10)\times 10^{-2}$
& $(-0.97\pm 0.01\pm 0.10)\times 10^{-2}$
\\
$C_{10}$ & $(2.12\pm 0.19\pm 0.24)\times 10^{-3}$
& $(1.95\pm 0.17\pm 0.24)\times 10^{-3}$
& $(1.86\pm 0.16\pm 0.23)\times 10^{-3}$
\\
\hline\end{tabular}
\caption[]{\it Wilson coefficients of the $\DSone$ effective Hamiltonian
at $\mu=1.5,2,3,4.5$ GeV. For $\mu<m_b$, the relation (\ref{eq:q10}) has been
used to reduce the operator basis.
We take $\Lambda_{QCD}^{(4)}=(330\pm 100)$ MeV and $m_t=(174\pm 17)$ GeV.
The values of the coefficients shown here correspond to the central values of
these parameters. The first error is due to the uncertainty on $\Lambda_{QCD}$,
the second is due to $m_t$.}
\label{tab:coeff}
\end{center}
\end{table}
}
\section{Relevant formulae}
\label{sec:formule}

In order to estimate $\epse$, we have to constrain the CP-violating
phase $\delta$ in the CKM matrix, by using the available experimental
information. To this end, we consider the CP-violating term in the
$K^0$--$\bar K^0$ mixing amplitude and the CP-conserving term for
$B^0$--$\bar B^0$ mixing.
In the following, we present all the formulae used in our analysis,
namely the expressions of $\ep$, $x_d$ and $\epse$ from the
$\DStwo$, $\DBtwo$, $\DSone$ effective Hamiltonian, respectively.

The effective Hamiltonian governing the $\DStwo$ amplitude is given by
\be
{\cal H}_{eff}^{\DStwo}
= \frac{G_{ F}^2}{16{\pi}^2}M_{ W}^2
({\bar s}{\gamma}^{\mu}_{ L}d)^2\left\{{\lambda}_c^2F(x_c)+{\lambda}_t
^2F(x_t)+2{\lambda}_c{\lambda}_tF(x_c,x_t)\right\}\, ,
\label{eq:eff_ham1}
\ee
where $x_q={m_q^2}/{M_{ W}^2}~$ and
the functions $F(x_i)$ and $F(x_i,x_j)$ are the so-called
{\it Inami--Lim} functions \cite{inami}, including QCD corrections
\cite{bur0}; $F(x_t)$ is known at the
next-to-leading order and has been included in our calculation.
{}From eqs. (\ref{eq:epsdef1}) and (\ref{eq:eff_ham1}),
one can derive the CP-violation parameter
\bea
\vert\epsilon\vert_{\xi=0}&=&C_{ \epsilon}B_{ K}A^2\lambda^6\sigma\sin\delta
\left\{F(x_c,x_t)+\right.\nn\\
& &\left.F(x_t)[A^2\lambda^4(1-\sigma\cos\delta)]-F(x_c)\right\}\, ,
\label{eq:epsilon_csizero_1}
\eea
where
\be
C_{ \epsilon}=\frac
{G_{ F}^2f_{ K}^2 M_{ K}M_{ W}^2}{6\sqrt 2{\pi}^2\Delta M}\, .
\ee
In eq. (\ref{eq:epsilon_csizero_1}), $\rho=\sigma
\cos \delta$, $\eta=\sigma \sin\delta$ and $\lambda$, $A$, $\rho$ and $\eta$
are the
parameters of the CKM matrix in the Wolfenstein parametrization \cite{wolf}.
$B_K$ is the renormalization group invariant $B$-factor, the
definition of which at the leading order is
\be
\langle \bar K\vert\left(\bar s\gamma^\mu_L d\right)^2\vert K\rangle =
\frac{8}{3}f_K^2M_K^2\alphas(\mu)^{6/25}B_K\, .
\ee

The $\DBtwo$ effective Hamiltonian is given by
\be
{\cal H}_{eff}^{\DBtwo}
= \frac{G_{ F}^2}{16{\pi}^2}M_{ W}^2 {\lambda}_t^2
({\bar b}{\gamma}^{\mu}_{ L}d)^2
F(x_t)\, .
\label{eq:eff_ham2}
\ee
Here $\lambda_t = V_{td} V^*_{tb}$.
{}From eq. (\ref{eq:eff_ham2}), one finds the $B^0$--$\bar B^0$ mixing
parameter
\bea
x_d &=&\frac{\Delta M_B}{\Gamma}=C_B \frac{\tau_B f_B^2}{M_B} B_B
A^2 \lambda^6 \Bigl( 1 +\sigma^2 -2 \sigma \cos\delta \Bigr)
F(x_t), \nn \\
C_B &=& \frac{G_F^2 M_W^2 M_B^2}{6 \pi^2}\, ,
\eea
where $B_B$ is the $B$-parameter relevant for $B-\bar B$ mixing,
the definition of which is analogous to the $B_K$ one.

We can write $\epsilon^{\prime}$ as
\be  \epsilon^{\prime}=i\frac{e^{ i(\delta_2-\delta_0)}}{\sqrt{2}}\frac{\omega}
{\mbox{Re}A_{ 0}}\left[\omega^{ -1}
(\mbox{Im}A_{ 2})^{\prime}-(1-\Omega_{ IB})\,\mbox{Im}A_{ 0}
\right]\, .
\ee
With respect to eq. (\ref{eq:epspdef}), we have here explicitly written
the isospin-breaking contribution $\Omega_{IB}$, see for example ref.
\cite{bur5},
\be
(\mbox{Im}A_{ 2})^{\prime}=(\mbox{Im}A_{ 2})-\Omega_{ IB}
(\omega\,\mbox{Im}A_{ 0})\, .
\ee

To compute $\mbox{Im}A_0$ and $(\mbox{Im}A_2)^{\prime}$, we need the
hadronic matrix elements of the operators $Q_i$ between a kaon and two pions.
Usually they are given in terms of the so-called $B$-parameters:
\bea
\langle \pi\pi(I=0)\vert Q_i(\mu)\vert K\rangle &=&
B^{1/2}_i(\mu)\langle \pi\pi(I=0)\vert Q_i\vert K\rangle _{VIA}\nn\\
\langle \pi\pi(I=2)\vert Q_i(\mu)\vert K\rangle &=&
B^{3/2}_i(\mu)\langle \pi\pi(I=2)\vert Q_i\vert K\rangle _{VIA}\, ,
\eea
where the subscripts $VIA$ means that the matrix elements are computed
in the vacuum insertion approximation.
The relevant $VIA$ matrix elements can be expressed in terms
of three quantities
\bea
X\!&=&\!f_{\pi}\left(M_{ K}^{ 2}-M_{\pi}^{ 2}\right), \\
Y\!&=&\!f_{\pi}\left(\frac{M_{ K}^{ 2}}{m_s(\mu)+m_d(\mu)}\right)
^2\sim 12\,X\left(\frac{0.15 \, \mbox{GeV}}{m_s(\mu)}\right)^2, \\
Z\!&=&\!4\left(\frac{f_{ K}}{f_{\pi}}-1\right)Y\, .
\eea
{}From $\Heff^{\DSone}$, the expressions of $(\mbox{Im}A_{ 2})^{\prime}$ and
$\mbox{Im}A_{ 0}$ in terms of Wilson coefficients and of the $B$-parameters
are obtained
\bea
\mbox{Im}A_{ 0} &=&-\GF \mbox{Im}\Bigl({ V}_{ ts}^{ *}{ V}_{ td}\Bigr)
\left\{-\left(C_{ 6}B_{ 6}+\frac{1}{3}C_{ 5}B_{ 5}\right)Z
+\left(C_{ 4}B_{ 4}+\right.\right.\nn\\
& &\!\left.\frac{1}{3}C_{ 3}B_{ 3}\right)X+
C_{ 7}B_{ 7}^{ 1/2}\left(\frac{2Y}{3}+\frac{Z}{6}+
\frac{X}{2}\right)+C_{ 8}B_{ 8}^{ 1/2}\left(2Y+\right.\nn\\
& &\!\left.\left.\frac{Z}{2}+\frac{X}{6}\right)-
C_{ 9}B_{ 9}^{ 1/2}\frac{X}{3}+\left(\frac{C_{ 1}
B_{ 1}^{ c}}{3}+C_{ 2}B_{ 2}^{ c}\right)X\right\}\, ,
\eea
\bea
(\mbox{Im}A_{ 2})^{\prime}\!&=&\!-G_{ F}\mbox{Im}\Bigl({ V}_{ ts}^{ *}{ V}_{
td}\Bigr)
\left\{C_{ 7}B_{ 7}^{ 3/2}\left(\frac{Y}{3}-\frac{X}{2}\right)+
\right. \nn \\
& &\!\left.C_{ 8}B_{ 8}^{ 3/2}\left(Y-\frac{X}{6}\right)+
C_{ 9}B_{ 9}^{ 3/2}\frac{2X}{3}\right\}\, .
\eea
Notice that the matrix elements of the electromagnetic left--right operators
$Q_{7,8}$, which belong to the $(8_L, 8_R)$ representation of
$SU(3)_L\otimes SU(3)_R$, contain $Y$ and do not vanish in the chiral limit.

The evaluation of the $B$-factors requires a non-perturbative technique.
The Wilson coefficients and the hadronic matrix elements both
depend on the regularization scheme.
In order to cancel this dependence (up to $O(\as^2)$), it is necessary
to control the matching between the $B$-parameters and the coefficients at
the next-to-leading order.
Notice that many non-perturbative methods (e.g. $1/N$ expansion) do not fulfil
this requirement.

Two different approaches to the matrix element evaluation have been used
in recent next-to-leading $\epse$ analyses:
\begin{itemize}
\item In our previous analysis \cite{ciuc1,epenew},
the numerical values of the $B$-parameters have been taken from lattice
calculations \cite{lattice}.
Suitable renormalization factors are introduced to take into account the
difference between the HV, NDR and  lattice regularization
schemes. For those $B$-parameters not yet computed on the
lattice\footnote{Indeed $B$-parameters,which give the main contribution
to the value of $\epse$, namely $B_6$, $B_8^{(3/2)}$ and $B_9^{(3/2)}$,
have already been computed on the lattice, see table \ref{tab:bpar}.},
we have made educated guesses, which are discussed in detail in ref.
\cite{epenew}.
\item A phenomenological approach has been implemented in ref. \cite{burasepe},
where
the $B$-parameters are constrained by using the experimental
information
from  CP-conserving processes, by assuming $SU(3)$ flavour symmetry and
deducing some constraints
relating hadronic matrix elements at the charm threshold.
Unfortunately, there is no way to determine the
most important
$B$-factors necessary  to estimate $\epse$, namely $B_6$ and $B_8$, which
remain essentially unconstrained in this approach.
\end{itemize}

\section{Results}
\label{sec:results}

In this section, the main results of our analysis are summarized.
These results have been obtained by varying the experimental quantities,
e.g. the value of the top mass $m_t$, $\tau_B$,  etc., and
the theoretical parameters, e.g. the  $B$-parameters,
the strange quark mass $m_s$, etc., according to their errors.
Values and errors of the input quantities used in the following are reported
in tables \ref{tab:val}--\ref{tab:bpar}.
We assume a Gaussian distribution
for the experimental quantities and a flat distribution
(with a width of 2$\sigma$) for the theoretical ones.
The only exception is $m_s$, taken from quenched lattice $QCD$ calculations,
for which we have assumed a Gaussian distribution, according to the results
of ref. \cite{ms}.
\par
The theoretical predictions ($\cos \delta$, $\epse$, etc.)
depend on several fluctuating parameters. We have obtained their
distributions  numerically,
from which we have calculated the central values and the errors
reported below.
\begin{table}[t]
\begin{center}
\begin{tabular}{|c|c|}
\hline\hline
Parameters & Values \\ \hline
$m_t$ & $(174 \pm 17)$ GeV \\
$m_s$(2 GeV) & ($128\pm 18$) MeV \\
$\Lambda_{QCD}^{n_f=5}$ & ($230 \pm 80$) MeV \\
$V_{cb}=A\lambda^2$ & $0.040\pm 0.006$ \\
$\vert V_{ub}/V_{cb}\vert=\lambda\sigma$ & $0.080\pm 0.015$ \\
$\tau_B$ & $(1.49\pm 0.12)\times 10^{-12}$ sec \\
$x_d$ & $0.685\pm 0.076$ \\
$(f_B B_B^{1/2})_{th}$ & $(200\pm 40)$ MeV \\
$\Omega_{IB}$ & $0.25\pm 0.10$ \\ \hline\hline
\end{tabular}
\caption[]{ \it{Values of the fluctuating parameters used in the numerical
analysis.}}
\label{tab:val}
\end{center}
\end{table}

\begin{table}[t]
\begin{center}
\begin{tabular}{|c|c|}
\hline\hline
Constants & Values \\ \hline

$G_F$ & $1.16634\times 10^{-5}\mbox{GeV}^{-2}$ \\
$m_c$ & 1.5 GeV \\
$m_b$ & 4.5 GeV\\
$M_W$ & 80.6 GeV \\
$M_{\pi}$ & 140 MeV \\
$M_K$ & 490 MeV \\
$M_B$ & 5.278 GeV \\
$\Delta M_K$ & $3.521\times 10^{-12}$ MeV \\
$f_{\pi}$ & 132 MeV \\
$f_K$ & 160 MeV \\
$\lambda=\sin\theta_c$ & 0.221 \\
$\ep_{exp}$ & $2.268\times 10^{-3}$ \\
$\mbox{Re}A_0$ & $2.7\times 10^{-7}$ GeV \\
$\omega$ & 0.045 \\
$\mu$ & $2$ GeV\\ \hline\hline
\end{tabular}
\caption[]{ \it{Constants used in the numerical analysis. }}
\label{tab:const}
\end{center}
\end{table}
\begin{table}
\begin{center}
\begin{tabular}{|c|c|c|c|c|c|c|}\hline\hline
$B_K^{rgi}$ & $B_9^{(3/2)}$ &  $B_{1-2}^{ c}$ &
 $B_{3,4}$ &
$B_{5,6}$ & $B_{7-8-9}^{(1/2)}$ & $B_{7-8}^{(3/2)}$\\
\hline
$0.75\pm 0.15$ & $0.62\pm 0.10$  &  $0 - 0.15^{(*)}$ & $1 -  6^{(*)}$ &
$1.0\pm 0.2$ & $1^{(*)}$ & $1.0\pm0.2$
\\ \hline\hline
\end{tabular}
\caption[]{{\it Values of the $B$-parameters, for operators renormalized at the
scale $\mu=2$ GeV. The only exception is $B_K^{rgi}$, which is the
renormalization group invariant $B$-parameter. $B_9^{3/2}$
has been taken equal to $B_K$, at any renormalization scale. The value reported
in the table is $B_9^{3/2}(\mu=2$ GeV).
Entries with a $^{(*)}$
are educated guesses, the others are taken from lattice QCD calculations.}}
\label{tab:bpar}
\end{center}
\end{table}

Using the values given in the tables and the formulae given in the
previous sections, we have obtained the following
results:
\begin{itemize}
\item[a)]
The distribution for $\cos \delta$, obtained by comparing
the experimental value of $\ep$ with its theoretical
prediction, is given in fig. \ref{fig:cd}. As already noticed in
refs. \cite{reina,ciuc1} and \cite{alig,schubert},
large values of $f_B$ and $m_t$ favour $\cos \delta > 0$,
given the current measurement of $x_d$. When the condition $160$ MeV
$\le f_B B_B^{1/2}\le 240$ MeV is imposed ($f_B$-cut),
most of the negative solutions disappear,
giving the dashed histogram of fig. \ref{fig:cd}, from which we estimate
\be
\cos \delta= 0.47 \pm 0.32\,\, .
\ee
\begin{figure}   
\begin{center}
\epsfxsize=1.08\textwidth
\leavevmode\epsffile{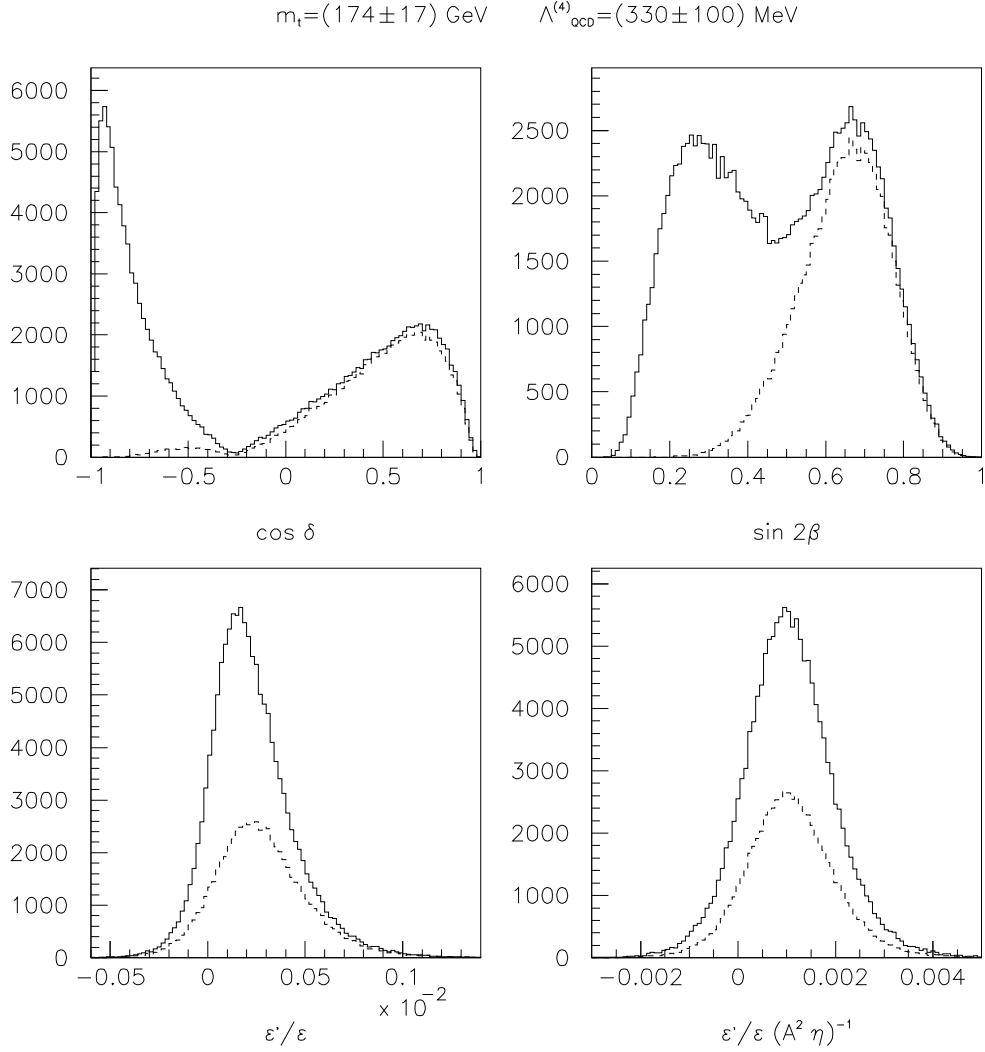}
\caption[]{\it{ Distributions of values for $\cos \delta$,
$\sin 2 \beta$, $\epse$ and $\epse (A^2\eta)^{-1}$,
for $m_t=(174 \pm 17)$ GeV,
using the values of the parameters given in tabs.
\ref{tab:val}--\ref{tab:bpar}. The solid histograms are obtained without
using the information coming from  $B_d$--$\bar B_d$ mixing. The dashed ones
use the $x_d$ information, assuming that
$160$ MeV $\le f_B B_B^{1/2}\le 240$ MeV. }}
\label{fig:cd}
\end{center}
\end{figure}
\begin{figure}   
\begin{center}
\epsfxsize=1.08\textwidth
\leavevmode\epsffile{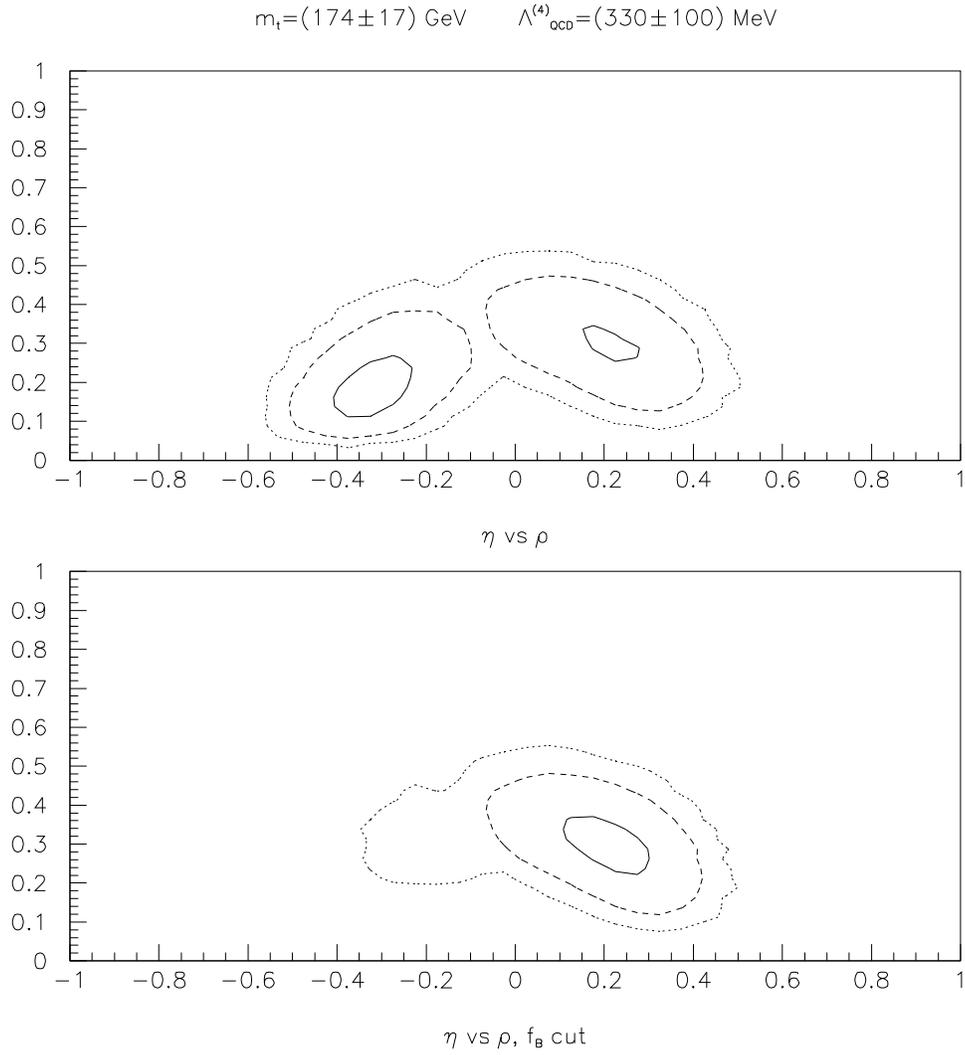}
\caption[]{\it{ Contour plots in the $\rho$--$\eta$ plane.
The solid, dashed and dotted contours contain
$5 \%$, $68 \%$ and $95 \%$ of the generated events respectively.
The contours are given by excluding or including the
$f_B$-cut. Similar results can be found in refs. \cite{alig,schubert}.}}
\label{fig:rhoeta}
\end{center}
\end{figure}
\begin{figure}   
\begin{center}
\epsfxsize=1.08\textwidth
\leavevmode\epsffile{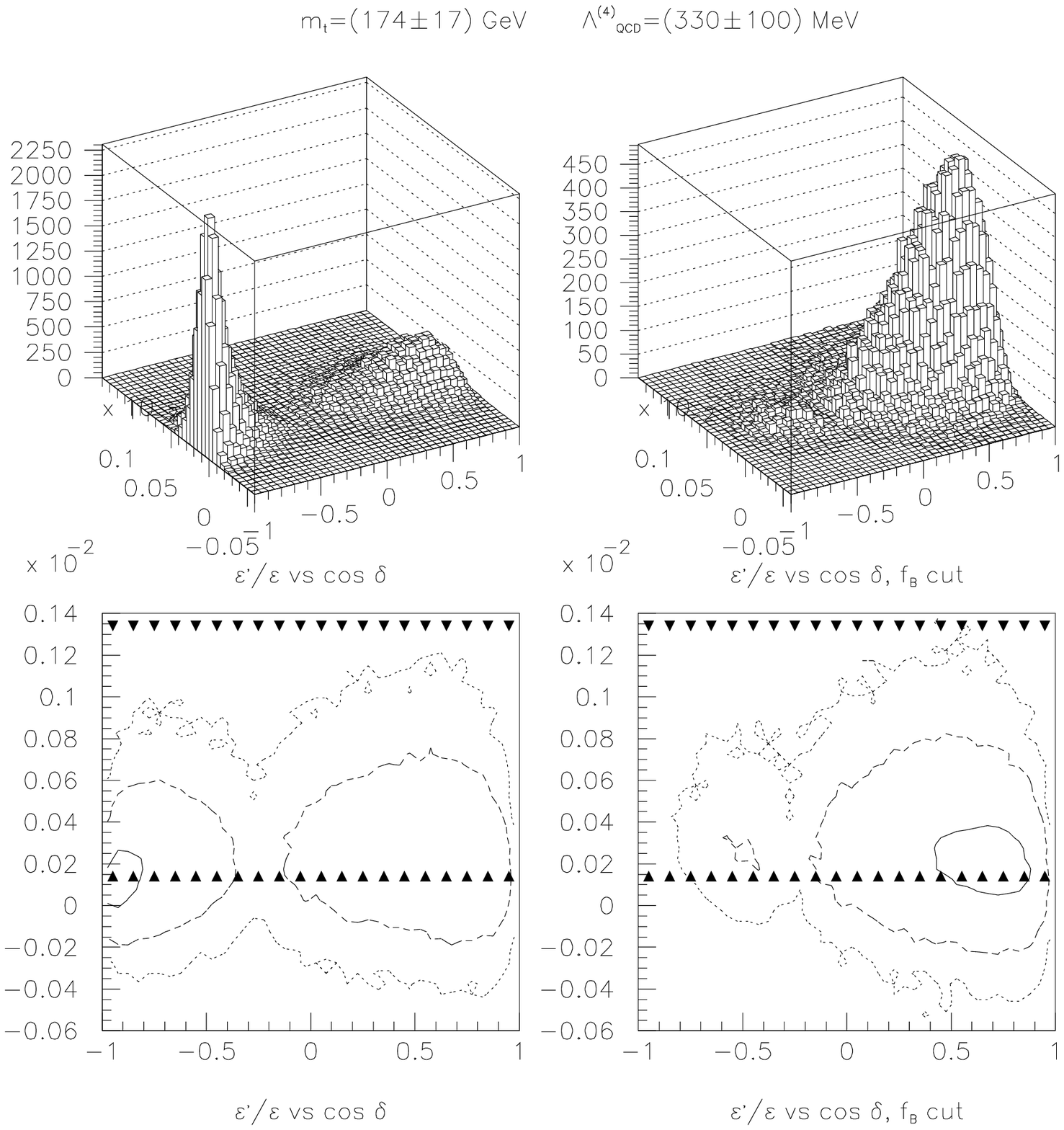}
\caption[]{\it{Distributions of the events in the
plane $\epse$--$\cos \delta$ without
and with the $f_B$-cut. The corresponding
contour plots are displayed below the Lego plots.}}
\label{fig:tuttoepe}
\end{center}
\end{figure}
\item[b)] A contour plot in the $\rho$--$\eta$ plane is
given in fig. \ref{fig:rhoeta}. It shows the current limits on the
unitarity triangle defined in fig. \ref{fig:triangle}.
\item[c)] In fig. \ref{fig:tuttoepe}, several pieces of information
on $\epse$ are provided. Lego plots of the distribution of the generated
events in the $\epse$--$\cos \delta$ plane are shown, without
and with the $f_B$-cut. In the same figure, the corresponding
contour plots are displayed. One notices a very mild dependence
of $\epse$ on $\cos \delta$. As a consequence, one obtains approximately
the same prediction in the two cases (see also fig. \ref{fig:cd}).
In the HV scheme the results are
\be
\epse = (2.3 \pm 2.1 )\times  10^{-4} \,\,\, {\rm no-cut}
\ee
and
\be
\epse = (2.8 \pm 2.4 )\times  10^{-4} \,\,\, f_B-{\rm cut}\, ,
\label{eq:epphv}
\ee
whereas in the NDR scheme we obtain
\be
\epse = (2.8 \pm 2.2 )\times  10^{-4} \,\,\, {\rm no-cut}
\ee
and
\be
\epse = (3.4 \pm 2.5 )\times  10^{-4} \,\,\, f_B-{\rm cut}\, .
\label{eq:eppndr}
\ee
By averaging the results given in eqs. (\ref{eq:epphv}) and
(\ref{eq:eppndr}), we obtain our best estimate
\be
\epse = (3.1\pm 2.5\pm 0.3)\times 10^{-4}\,\,\, f_B-{\rm cut}\, ,
\ee
where the third error comes from the difference of the central values
in the two schemes and gives an estimate of the uncertainty due to
higher-order corrections.
\end{itemize}
\par
A similar result has been obtained in ref. \cite{burasepe}, using a
different approach to the hadronic-matrix-element evaluation. They quote
\be
\frac{\epp}{\epsilon}=(6.7\pm 2.6)\times 10^{-4}
\ee
for $m_t=130$ GeV. For this value of the top mass, the cancellation between
penguin and electropenguin contributions is less effective, thus their
$\epse$ prediction is significantly larger than ours. Actually the two
predictions agree, once the difference in the top mass is
taken into account\footnote{Our analysis gives $\epse=(6.3\pm 2.3)\times
10^{-4}$ for $m_t=130$ GeV.}.
It is reassuring that theoretical predictions, obtained by using quite
different approaches to matrix elements evaluation, are in good agreement.

On the basis of the latest analyses, it seems very difficult
that $\epse$ is larger than $10 \times 10^{-4}$.
Theoretically, this may happen by taking the matrix elements of the
dominant operators, $Q_6$ and $Q_8$, much more different than
it is usually assumed.
One possibility, discussed in ref. \cite{burasepe}, is to take
$B_6 \sim 2$ and $B_8 \sim 1$, instead of the usual values
$B_6 \sim B_8 \sim 1$.
To our knowledge, no coherent theoretical approach can accommodate
such large values of $B_6$.

\end{document}